\documentclass[twocolumn,showpacs,preprintnumbers,amsmath,amssymb]{revtex4}

\usepackage{csquotes}
\usepackage{amsmath}
\usepackage{amssymb}

\newcommand{\system}{Sr$_3$Ru$_2$O$_7$}

\title{Electronic nematicity and its relation to quantum criticality in \system\ studied by thermal expansion}

\titlerunning{Electronic nematicity in \system\ }

\author{%
  C. Stingl\textsuperscript{\textsf{\bfseries 1}},
  R.S. Perry\textsuperscript{\textsf{\bfseries 2}},
  Y. Maeno\textsuperscript{\textsf{\bfseries 3,4}},
   and
  P. Gegenwart\textsuperscript{\Ast,\textsf{\bfseries 1}}}

\authorrunning{C. Stingl et al.}

\mail{e-mail
  \textsf{Philipp.Gegenwart@phys.uni-goettingen.de}, Phone:
  +49-551-39-7607, Fax: +49-551-39-19546 }

\institute{%
  \textsuperscript{1}\, I. Physikalisches Institut, Georg-August-Universit\"{a}t
G\"{o}ttingen, Friedrich-Hund-Platz 1, 37077  G\"{o}ttingen,
Germany\\
\textsuperscript{2}\, Scottish Universities Physics Alliance, School of Physics, University of Edinburgh, Mayfield Road, Edinburgh EH9 3JZ, Scotland\\
\textsuperscript{3}\, International Innovation Center, Kyoto University, Kyoto 606-8501, Japan\\
\textsuperscript{4}\, Department of Physics, Kyoto University, Kyoto 606-8502, Japan\\
    }

\received{XXXX, revised XXXX, accepted XXXX} 
\published{XXXX} 


\abstract{%
\abstcol{We report high-resolution measurements of the in-plane thermal expansion anisotropy in the vicinity of the electronic nematic phase in \system\ down to very low temperatures and in varying magnetic field orientation. For fields applied along the c-direction, a clear second-order phase transition is found at the nematic phase, with critical behavior compatible with the two-dimensional Ising universality class (although this is not fully conclusive).} {Measurements in a slightly tilted magnetic field reveal a broken four-fold in-plane rotational symmetry, not only within the nematic phase, but extending towards slightly larger fields. We also analyze the universal scaling behavior expected for a metamagnetic quantum critical point, which is realized outside the nematic region. The contours of the magnetostriction suggest a relation between quantum criticality and the nematic phase.}
}

\begin{document}

\maketitle   

\section{Introduction}
The layered ruthenate \system\ offers the unique possibility to study the interrelation between two phenomena that in recent years have been of considerable interest in condensed matter physics, i.e. quantum criticality and electronic nematic order. Quantum critical points (QCPs) arise from the continuous transformation between different ground states by the variation of an external parameter such as pressure, magnetic field or chemical composition. Quantum criticality is particularly interesting in metals, since the charge carriers undergo anomalous scattering leading to deviation from Landau Fermi liquid behavior. It has been found in many different material classes, including heavy-fermion metals~\cite{gegenwart08}, iron pnictide~\cite{Matsuda} and cuprate superconductors~\cite{broun08}. In \system, quantum criticality results from the suppression of the critical temperature of a first-order metamagnetic transition to absolute zero temperature, as the field angle is rotated towards the $c$-axis. Pronounced non-Fermi liquid effects have been detected in measurements of the electrical resistivity~\cite{grigera01}, thermal expansion ~\cite{gegenwart06}, and entropy~\cite{rost09}.

The second interesting effect in \system\ is an unusual electronic state which is characterized by a broken rotational symmetry, called electronic nematic order. Related behavior has also been found in cuprate and iron-pnictide superconductors~\cite{daou10,chu10} as well as two-dimensional (2D) electron gases~\cite{lilly99}. In \system, the first experimental evidence for electronic nematicity was a striking in-plane anisotropy of the electrical resistivity~\cite{borzi07}. Since this nematic phase occurs precisely in the vicinity of the QCP, the two phenomena appear to be fundamentally linked.

\system\ crystallizes in the $Bbcb$ structure. While the symmetry is lowered from tetragonal to orthorhombic by a cooperative rotation of RuO octahedra, the $a$ and $b$-axis have almost equal length so that the system can be viewed as pseudotetragonal. Transport is strongly two-dimensional and occurs in the RuO bilayers. The ground state of \system\ is an exchange enhanced paramagnet. In an external magnetic field applied parallel to the $c$-axis, it undergoes a series of metamagnetic transitions, two of which are first order at low temperatures. For $T\rightarrow0$, these two first-order transitions occur at $B_\text{c1}=7.8$\,T and $B_\text{c2}=8.1$\,T~\cite{grigera04}. At $T\lesssim 1$\,K the electronic nematic state is bounded in field by these two transitions. For a recent comprehensive review on the material, see~\cite{mackenzie12}.

While elastic neutron scattering experiments could not detect a structural change in connection with the formation of the nematic phase~\cite{borzi07}, we have recently resolved a very small ($10^{-6}$) but symmetry-breaking distortion as the phase is entered~\cite{stingl11}. This lattice effect seems to be driven by the electronic nematic state. In this paper, we will explore a wider region of the phase diagram by means of thermal expansion measurements.

\section{Experiment}

\begin{figure}\centering
\includegraphics[width=\columnwidth]{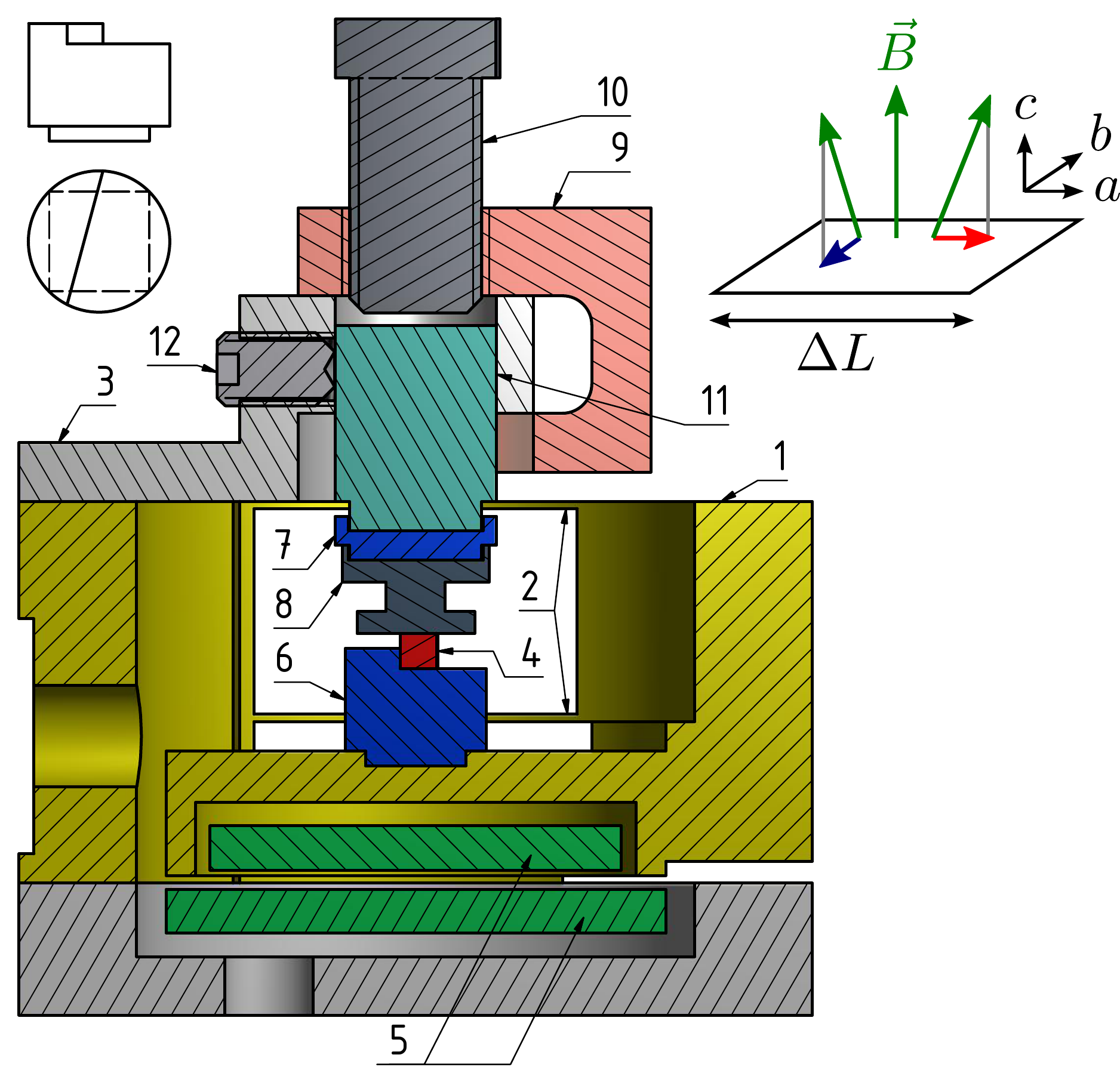}
\caption{Dilatometer with thermally decoupled sample. (1)~movable frame, (2)~flat springs, (3)~fixed frame, (4)~sample, (5)~capacitor plates,
(6)~sample holder and (7)~thermal insulator, both made from graphite, (8)~thermal anchor, (9)~removable fastening device, (10)~adjustment screw, (11)~piston, (12)~locking screw. Top left: sample holder for  a rotation of $\Phi=15^{\circ}$ around the vertical axis. Top right: sketch of the experimental configuration for $ \Delta L \parallel B_\text{in-plane}$ ({\color{red}red}) and $\Delta L \perp B_\text{in-plane}$ ({\color{blue}blue}).}
\label{dilatometer}
\end{figure}

For our work, we employ the miniaturized capacitive dilatometer described in \cite{kuechler12}, to which we have made a number of modifications, cf. Fig.~\ref{dilatometer}. The adjustment screw is here replaced by a sliding piston, which is clamped from the side by a locking screw. The initial capacitance is set with an external, removable adjustment screw. This design further reduces the vertical height of the setup and makes the capacitance more stable against mechanical shock which can not always be avoided during assembly of the cryostat.

In order to improve thermal equilibrium at the lowest temperatures, the sample is thermally decoupled from the dilatometer by insulator parts made from POCO AXM-5Q graphite~\cite{woodcraft03} and thermally anchored by a silver cylinder\footnote{A conducting sample as the one in this work is best cooled via direct metallic contact. For insulating samples, it has proven helpful to apply a small amount of Apiezon N grease between the sample and the silver piece and then warm the dilatometer slightly with a heat gun so that excess grease can flow out.} which is directly connected to the temperature stage with a silver loom. The bottom graphite piece plugs into the dilatometer base with a square end and doubles as a sample holder. The angle at which the edge is cut defines the rotation of the sample around the vertical axis.

The background effect resulting from the thermal expansion of the dilatometer itself was measured with all parts except the sample and subtracted from all shown curves. For fields between 7 an 8\,T, thermal expansion background is approx. $1\cdot10^{-8}\,\text{cmK}^{-1}$, which is 1\,--\,2 oders of magnitude smaller than the signal from the sample. The flat springs in our dilatometer exert a force of $\approx 3\,\text{N}$ on the sample, corresponding to a uniaxial pressure of 15\,bar on an area of $1.85\,\text{mm}^2$.

Measurements were performed on the same high-quality crystal investigated in previous studies~\cite{gegenwart06,stingl11}. The sample is mounted such that length changes are measured parallel to its pseudotetragonal $a$-axis, $\Delta L \parallel a$. The entire dilatometer is rotated by 90~degrees about the horizontal axis in Fig.~\ref{dilatometer} so that the magnetic field points along the $c$~direction, $B\parallel c$. This tunes \system\ to its QCP. A further small rotation of the dilatometer will then create an in-plane magnetic field component $B_\text{in-plane}\parallel \Delta L$, while slightly rotating the sample around the vertical axis in Fig.~\ref{dilatometer} yields $B_\text{in-plane}\perp \Delta L$. These two configurations are illustrated in the top right part of Fig.~\ref{dilatometer}.

\section{Nematic Phase Transition}

\begin{figure}
\includegraphics[width=\columnwidth]{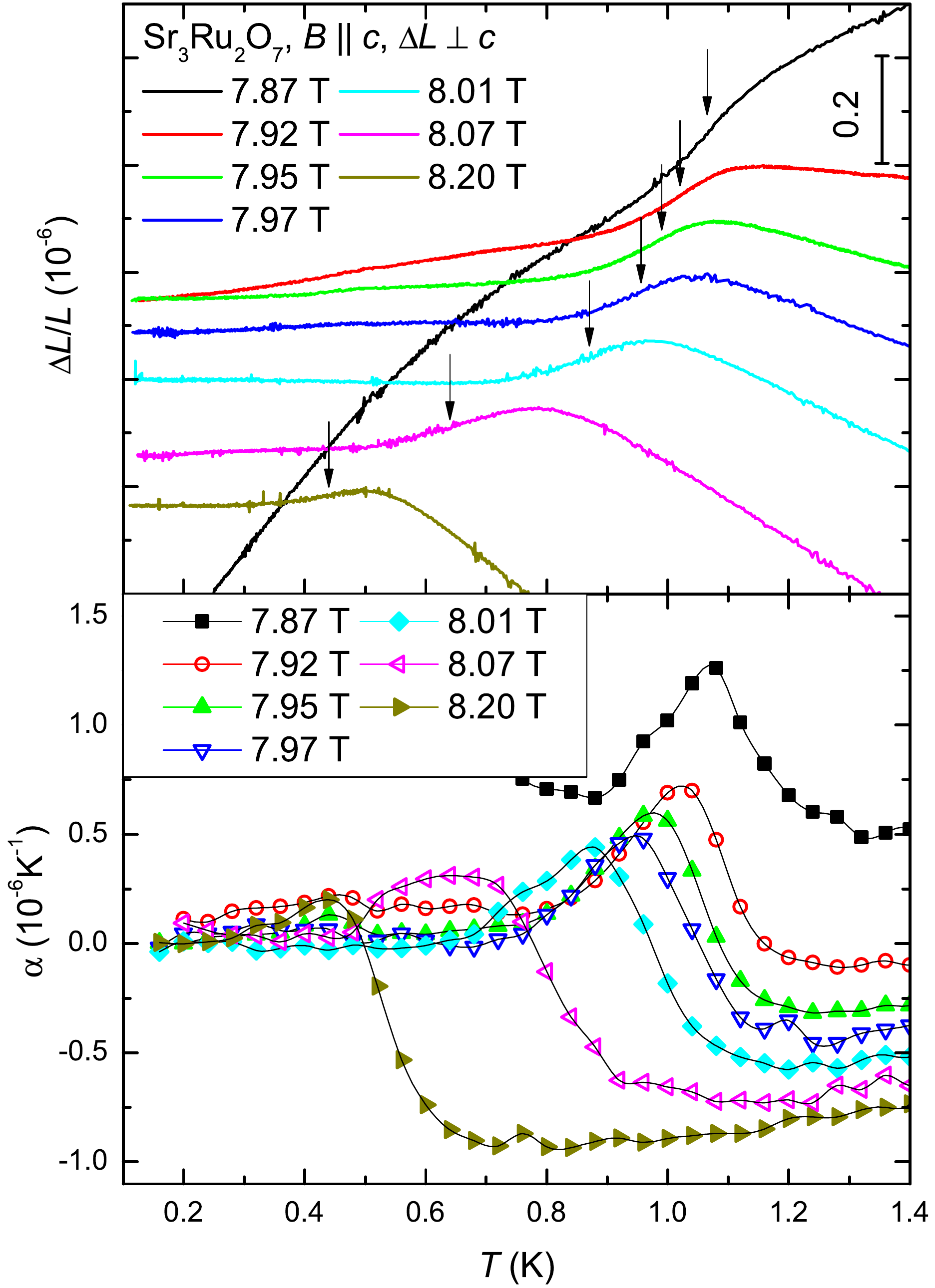}
\caption{Temperature dependence of the isofield length change (upper plot) and isofield thermal expansion (lower plot) of \system\ for various magnetic field applied along the $c$-direction. The arrows indicate the positions of the nematic phase transition as determined from the local maxima in the thermal expansion coefficient.}
\label{delta_l_inside}
\end{figure}

\begin{figure}
\includegraphics[width=\columnwidth]{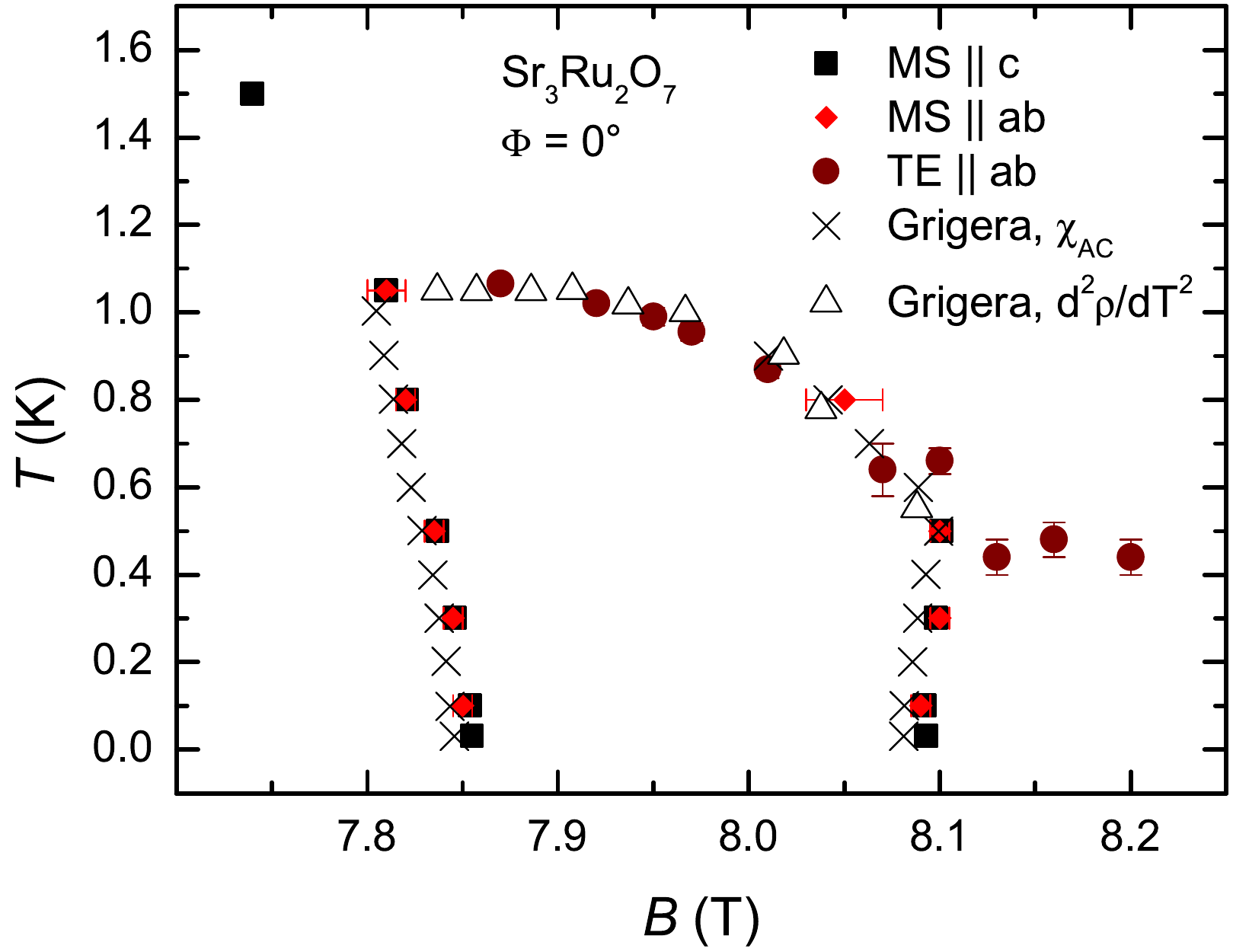}
\caption{Phase diagram of \system\ constructed from the loci of maxima in the $c$-axis (black squares) and in-plane (red diamonds) magnetostriction as well as the in-plane thermal expansion coefficient (circles). The phase boundaries agree well with data from~\cite{grigera04}.}
\label{pd_0deg}
\end{figure}

At first we will focus on length change measurements in the ab-plane with the field aligned parallel to the $c$-axis of the crystal. Figure~\ref{delta_l_inside} shows data for different magnetic fields between the first-order metamagnetic transitions. At approx. 1\,K, a clear spontaneous relative length change is observed, with a magnitude of about $1\cdot10^{-7}$. At a sample length of 1.5\,mm, this corresponds to $\Delta L$=1.5\,\AA. The thermal expansion coefficient $\alpha$ is obtained by local linear fits over intervals of 80\,mK. For all shown fields, $\alpha$ exhibits a step-like discontinuity with a superimposed peak and a FWHM of $\approx200$\,mK.

The phase diagram in Fig.~\ref{pd_0deg} is constructed by plotting the loci of maxima in $\alpha$ (arrows in Fig.~\ref{delta_l_inside}) together with magnetostriction data from \cite{stingl10}. The location of the phase boundary is in excellent agreement with data from transport and susceptibility experiments  \cite{grigera04}. This clearly shows that the signature in $\alpha$ is associated with entry into the nematic phase. At the first-order phase boundaries, a shift of approx. 10\,mT with respect to data points from other workers is apparent, which likely is caused by the uniaxial pressure which is necessarily exerted on the sample \cite{stingl11}.

The nematic phase has previously been found to be bounded in field by the two first-order metamagnetic transitions at $B_\text{c1}=7.8$\,T and $B_\text{c2}=8.1$\,T~\cite{borzi07,grigera04}. However, the signature in $\alpha$ is also observed for $B>8.1$\,T, indicating that the nematic regime might extend even beyond the second transition. This possibility will be discussed below.

\begin{figure}
\includegraphics[width=\columnwidth]{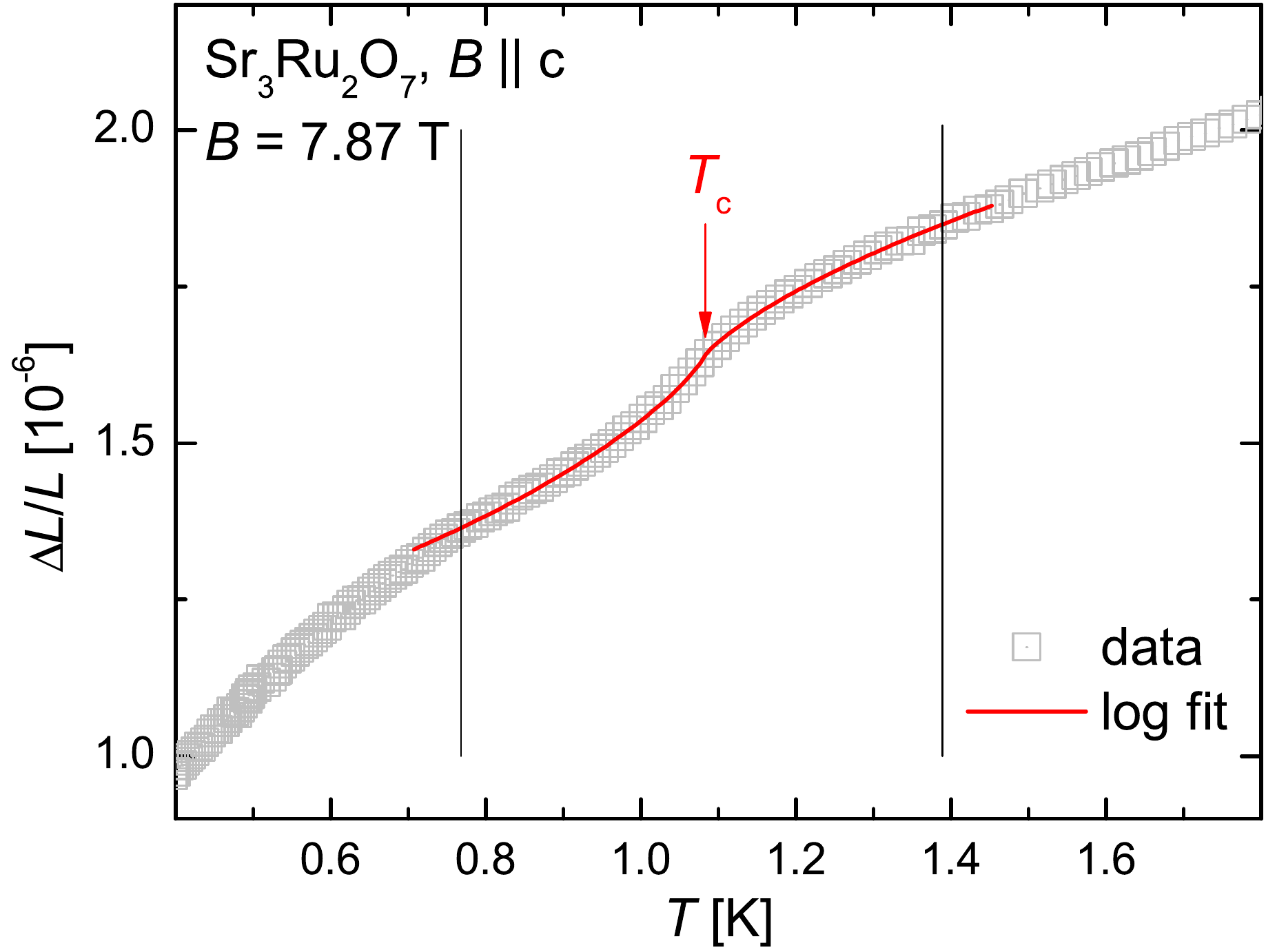}
\caption{Relative length change of \system\ at $B=7.87$~T applied along the $c$-axis. Within the temperature regime indicated by the two vertical lines the data have been described by a fit using eq. (1) according to critical behavior in the 2D Ising universality class (red line).}
\label{fit}
\end{figure}

\begin{table}\centering
\begin{tabular}{llll}
	parameter				&	value for log fit						& units				\\
	\hline
	$A^+$						& 0.215(8)			& $10^{-6}$\\
	$A^-$						&	0.310(6)			& $10^{-6}$\\
	$T_\text{c}$		&	1.083(3)			& K	\\
	$B$							&	1.640(4)			& $10^{-6}$\\
	$C$							&	0.254(15)			& $10^{-6}$\\
	\hline
\end{tabular}
\caption{Parameters for the description of the length change data shown in Fig.~\ref{fit} by eq. (1)}\label{params}
\end{table}

We now analyze the thermal-expansion signature at the nematic transition (Fig.~\ref{delta_l_inside}). Compatible with a second-order transition, a broadened step in $\alpha(T)$ is observed. Superposed to the step, we also observe a peak in $\alpha(T)$. Since we did not find any thermal hysteresis, there is no evidence for an incipient discontinuous transition. We therefore ascribe the peak in $\alpha(T)$ to critical fluctuations. It is thus interesting to investigate whether the observed behavior could be described by classical critical behavior of a known universality class. The latter should depend only on the dimensionality of the system and the degrees of freedom of the order parameter.

Since in \system, the important physics takes place in the RuO planes, which are separated along the $c$-axis by Sr atoms, the system is quasi-two-dimensional. From the observed symmetry-breaking lattice distortion~\cite{stingl11}, it follows that the phase near 8\,T is characterized by an Ising nematic order. Microscopically, this may be related to a partial orbital order~\cite{stingl11,lee10}. A straightforward assumption would then be that the system falls into the universality class of the 2D Ising model. For this model, a logarithmic divergence of the specific heat in the critical region is expected, $c_\text{cr}(T)\propto \log|t|$, with the reduced temperature $t=(T-T_\text{c})/T_\text{c}$. If the Gr\"{u}neisen law $\alpha\propto c$ holds, as expected for a classical (temperature driven) phase transition, the same critical behaviour should be found in the thermal expansion coefficient $\alpha$. Such a correspondence was confirmed e.g. in the 3D Heisenberg antiferromagnet EuTe~\cite{scheer92}.

In order to determine a critical exponent from thermal expansion, it is more accurate to directly fit to the relative length change $\Delta L/L$ data rather than its numerical derivative $\alpha=d(\Delta L/L)/dT $~\cite{krellner09}. The 2D Ising model predicts
$\alpha = -\left(A^+\Theta(t) + A^-\Theta(-t) \right)\log|t|$
where $\Theta$ denotes the Heaviside step function, $A^\pm$ are  the coefficients  above and below the critical temperature, and $t=T/T_\text{c}-1$ is the reduced temperature. Integrations yields
\begin{multline}\label{fit1}
	\frac{\Delta L}{L}(T) = 
	 -\left(A^+\Theta(t) - A^-\Theta(-t) \right) (|t|log|t|-|t|)\\
	+ B + Ct \quad ,
\end{multline}
where the last two terms allow for a linear background in the length change. This expression allows us to fit simultaneously the length change above and below the phase transition, and the critical temperature $T_\text{c}$ is included as a free parameter.

Considering the signature of the nematic phase transition in the thermal expansion $\Delta L(T)/L$ for different magnetic fields shown in Fig.~\ref{delta_l_inside}, it is seen that while all curves show a step-like increase of the sample length as the \enquote{roof} of the nematic phase is crossed, this signature is superimposed on different backgrounds. In most of the measured fields, the slope has different signs below and above the transitions. For our analysis, we concentrate on the data for $B=7.87$\,T, where the background is most uniform (the background slope is about $0.25\cdot10^{-6}$\,K$^{-2}$).
Figure~\ref{fit} shows the relative length change data at 7.87\,T together with a fit according to eq.~(\ref{fit1}). The resulting parameters are listed in Table~\ref{params}.

Clearly a nice description of the data by the 2D Ising model is possible. However, one should be careful, since the critical region for classical phase transitions is generally rather small, i.e. only of order $t\leq0.1$. The correct determination of critical exponents is therefore highly non-trivial and requires precise measurements at temperatures very close to the phase transition~\cite{wosnitza07}. In our case, the transition is quite broadened. Therefore, we need to extend the fitted temperature regime to $t\leq0.2$ in order to fully capture the phase transition. To get an impression on how reliable the logarithmic divergence could be differentiated from a weak power-law, we have also assumed a power-law divergence of the form $\alpha_\text{cr} = \left(A^+\Theta(t) + A^-\Theta(-t) \right)|t|^{-a}$ with a critical exponent~$a$, integration of which gives
\begin{multline}\label{fit2}
	\frac{\Delta L}{L}(T) = \left(A^+\Theta(t) - A^-\Theta(-t) \right) |t|^{-(a-1)} \\
		+ B + Ct	\quad.
\end{multline}
Leaving the exponent $a$ as free parameter, the fit (not shown) converges to very small $a$ values of the order of $10^{-3}$, consistent with the logarithmic divergence. However, we have also performed a fit with fixed, arbitrary value of the exponent, e.g. $a = 0.2$. Interestingly, eq.~(\ref{fit2}) is still a reasonable good fit (not shown) to the data with such a value for $a$. Thus, the signature in $\alpha$ is consistent with a logarithmic divergence but this is not conclusive, since a power law behaviour can not entirely be ruled out on the basis of our data.

\section{Lattice distortion}

\begin{figure}
\includegraphics[width=\columnwidth]{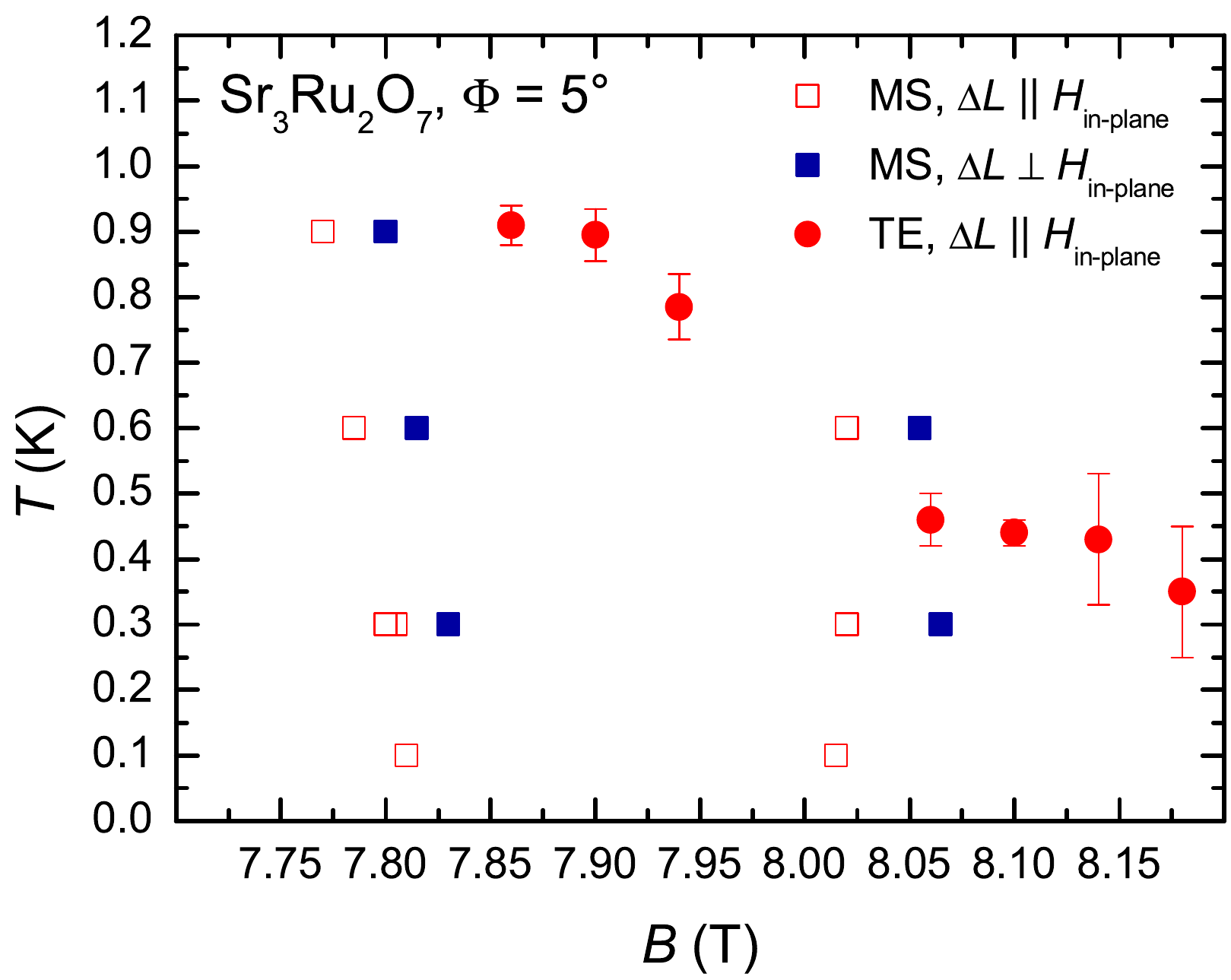}
\caption{Phase diagram for the magnetic field tilted from the $c$-axis by $\Theta=5^\circ$. The limited precision of rotation angle leads to slight differences in the two measured configurations.}
\label{pd_5deg}
\end{figure}

\begin{figure}
\includegraphics[width=\columnwidth]{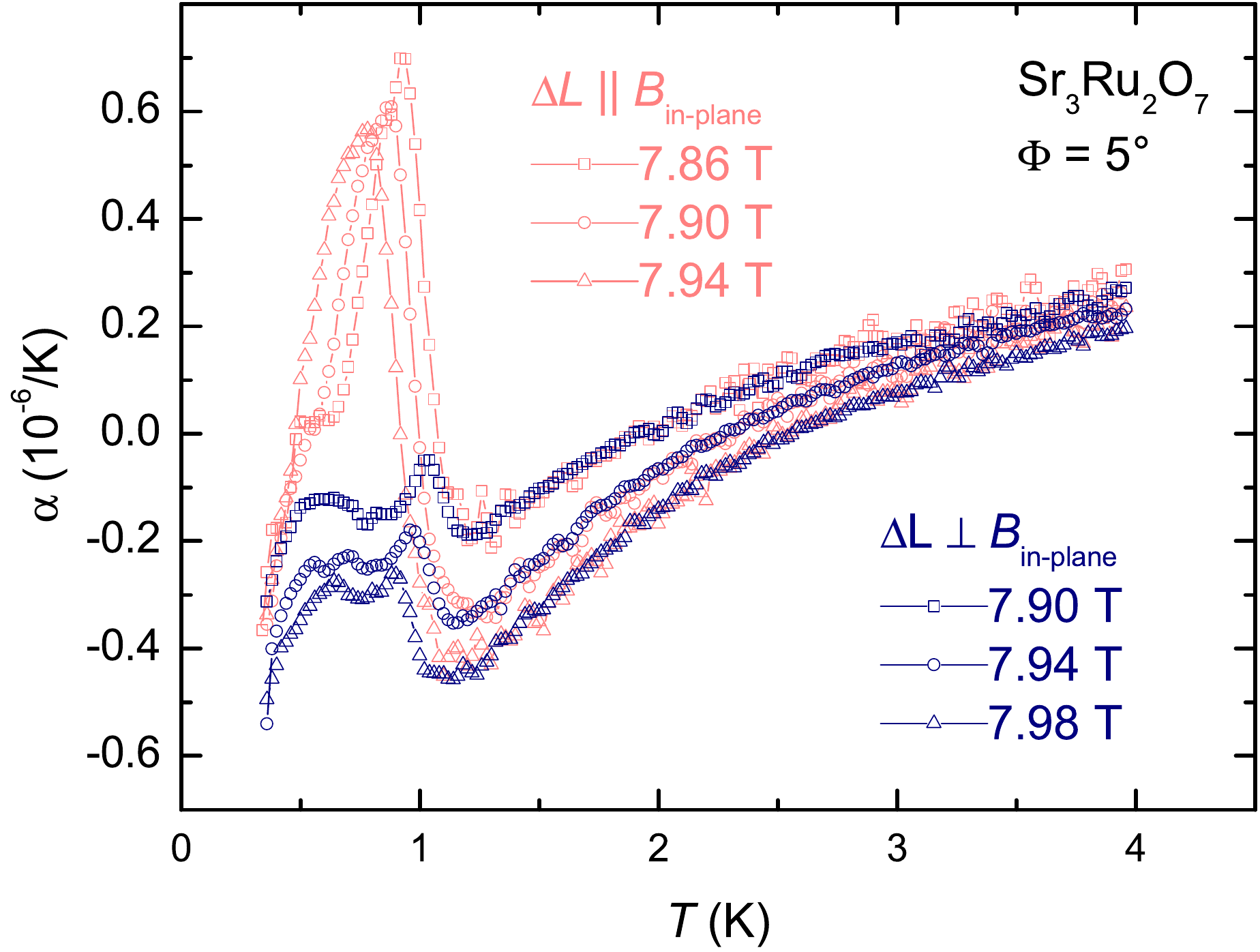}
\caption{Thermal expansion in fields tilted by $\Theta=5^\circ$ away from the $c$-axis. For fields in the nematic regime between 7.8 and 8.1 T, a symmetry breaking is observed below approx. 1\,K.}
\label{5deg-alpha_inside}
\end{figure}

\begin{figure}
\includegraphics[width=\columnwidth]{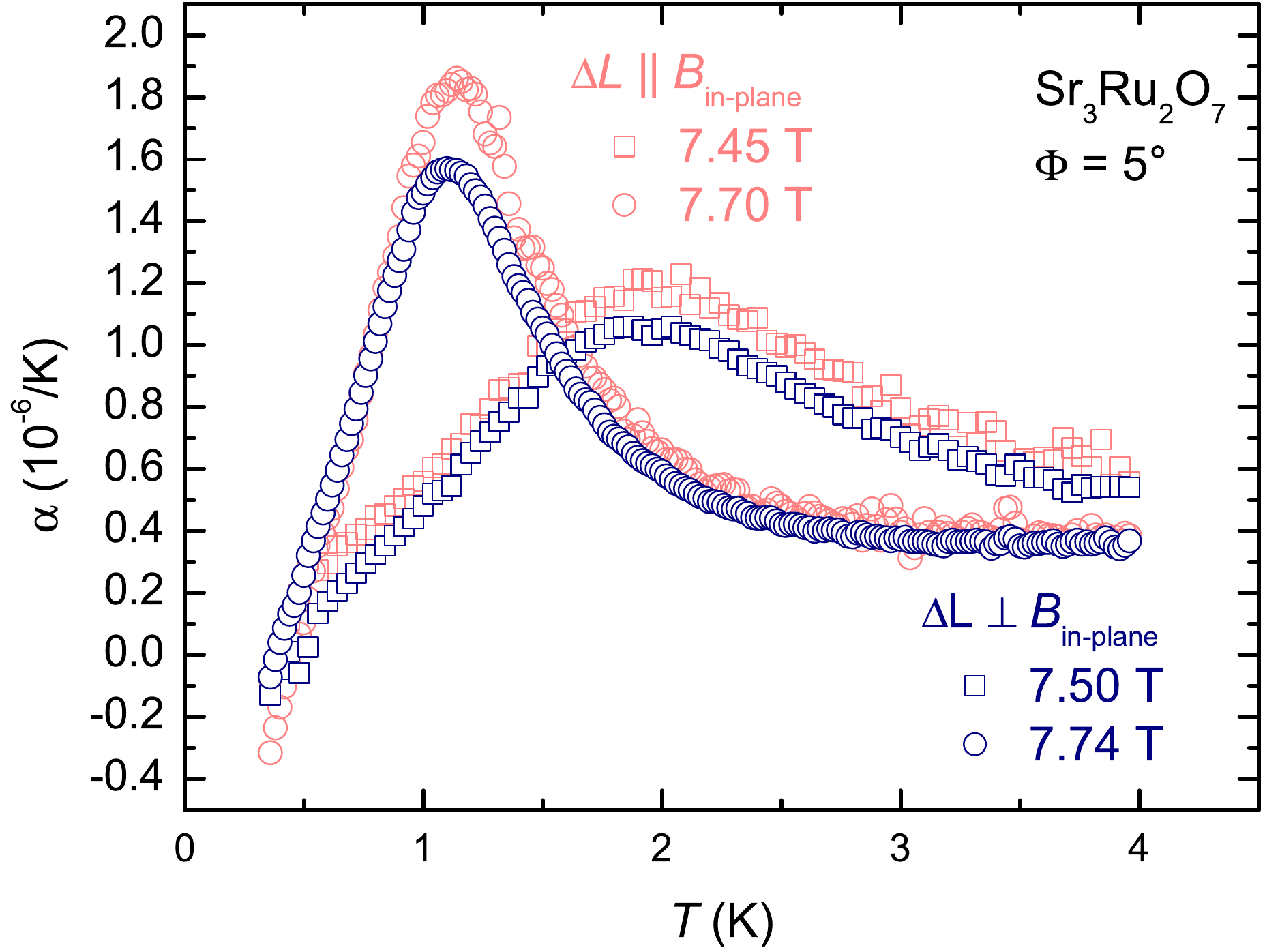}
\caption{Thermal expansion for $\Theta=5^\circ$ on the low-field side: In the region of the second-order metamagnetic transition at approx. 7.5 T, as well as for fields slightly below the onset of nematicity, no symmetry breaking occurs.}
\label{5deg-alpha_lowfield}
\end{figure}

\begin{figure}
\includegraphics[width=\columnwidth]{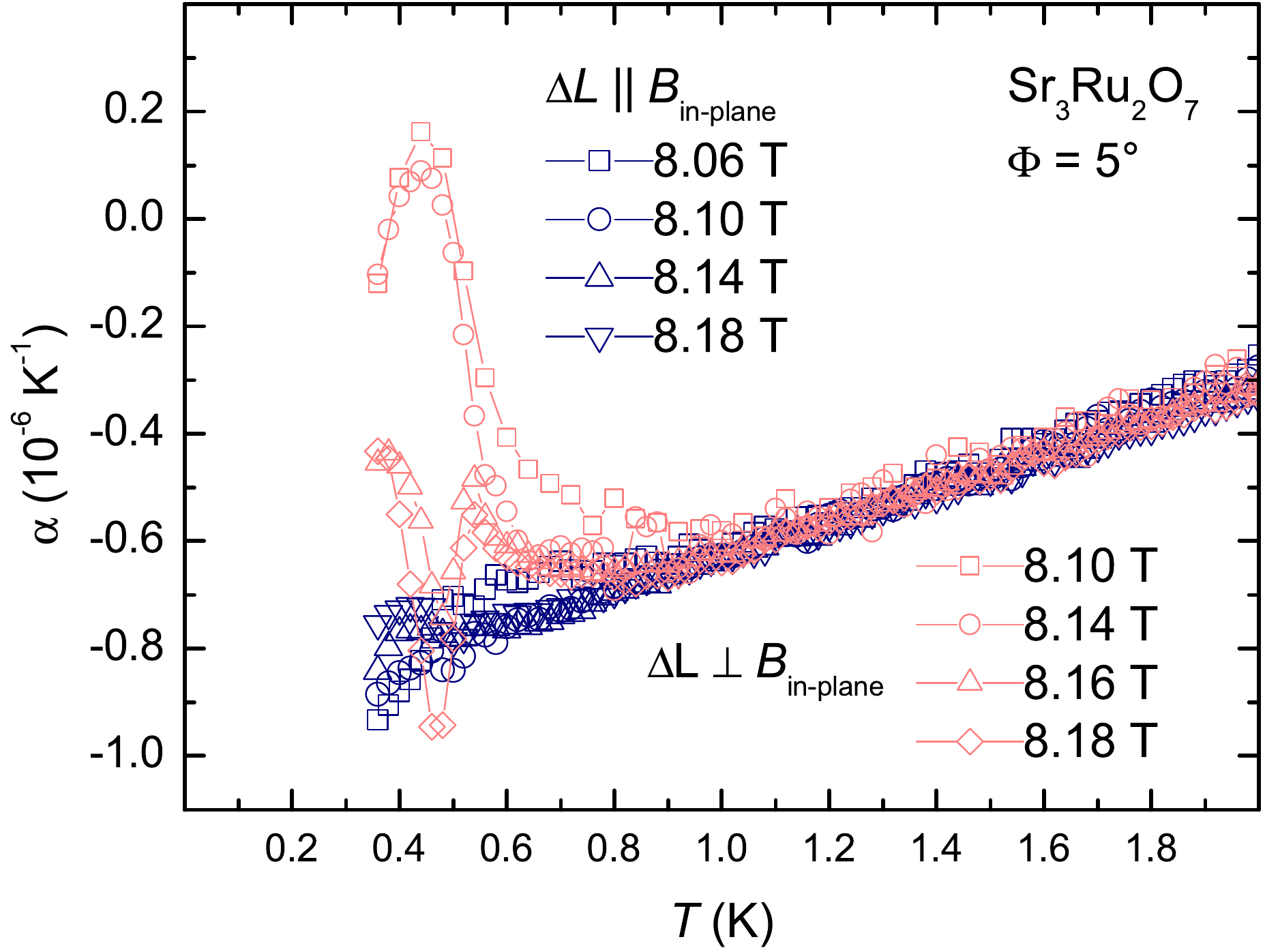}
\caption{Thermal expansion for $\Theta=5^\circ$ on the high-field side: the asymmetry persists for fields beyond the second first-order transition, but decreases in magnitude with increasing field.}
\label{5deg-alpha_highfield}
\end{figure}

We now turn to measurements with the magnetic field tilted away from the $c$-axis by an angle of $\Theta=5^\circ$. The data were taken in two separate runs, one for $\Delta L \parallel B_\text{in-plane}$ and another for $\Delta L \perp B_\text{in-plane}$. As the metamagnetic transitions and thereby the nematic phase shift to lower fields with increasing angle, isothermal magnetostriction has first been measured in order to determine the position of the two first-order metamagnetic phase transitions. We can only set the angle $\Theta$ with an accuracy of about 1-2 degrees and can not rotate the sample in situ. As can be seen from the phase diagram for $\Theta=5^\circ$ in Fig.~\ref{pd_5deg}, this leads to slightly different positions of the metamagnetic signatures for the two separate runs. In the following, we compensate for this slight angle error by comparing data for $\Delta L \parallel B_\text{in-plane}$ taken at a field $B$ with data for $\Delta L \perp B_\text{in-plane}$ taken at $B+40$\,mT.\footnote{In our earlier study~\cite{stingl11}, the error in the tilt angle was smaller and no correction was required.}

Figure~\ref{5deg-alpha_inside} shows the thermal expansion coefficient $\alpha$ for both directions as a function of temperature for three different magnetic fields in the nematic regime. For temperatures larger than 1\,K, the thermal expansion is perfectly isotropic within the basal plane. Here, the small in-plane field is not strong enough to break the fourfold symmetry. At lower temperatures however, the small in-plane component is sufficient to expose the intrinsic asymmetry.

In order to prove that the observed structural distortion can unambiguously be linked to the electronic nematic phase, we have performed the same measurements for smaller and larger fields outside of the nematic regime. Figure~\ref{5deg-alpha_lowfield} shows data for the low-field side. Even though the curves for the two perpendicular directions do not perfectly coincide (note that the absolute values of $\alpha$ are a factor of 5 larger than those in Fig.~\ref{5deg-alpha_inside}, so that the explicit symmetry breaking by the in-plane field might become visible here), the shape of $\alpha(T)$ is qualitatively the same and no distortion can be observed: Thermal expansion is indeed isotropic for all temperatures on the low-field side of the phase. This situation is however not so clear on the high-field side. Fig.~\ref{5deg-alpha_highfield} shows that a symmetry breaking can still be observed beyond the second first-order transition at~$B_\text{c2}$, although both the temperature of its onset and its magnitude decrease rapidly with increasing field. This is a further indication that the nematic regime extends even beyond~$B_\text{c2}$. The nematic order parameter does not suddenly vanish upon crossing the second first-order metamagnetic transition, but rather seems to decrease continuously to zero.

This observation is also supported by small details in previous resistivity measurements. As can be seen from Fig.~1b in~\cite{borzi07}, the resistive anisotropy sets in sharply at the first first-order transition. At the second first-order transition, however, the curve features a broad shoulder, indicating that the anisotropy only gradually reduces to zero. The origin of the anisotropy extending beyond $B_\text{c2}$ is currently under investigation~\cite{mackenzie12}.
\section{Quantum Criticality}

\begin{figure}
	\includegraphics[width=\columnwidth]{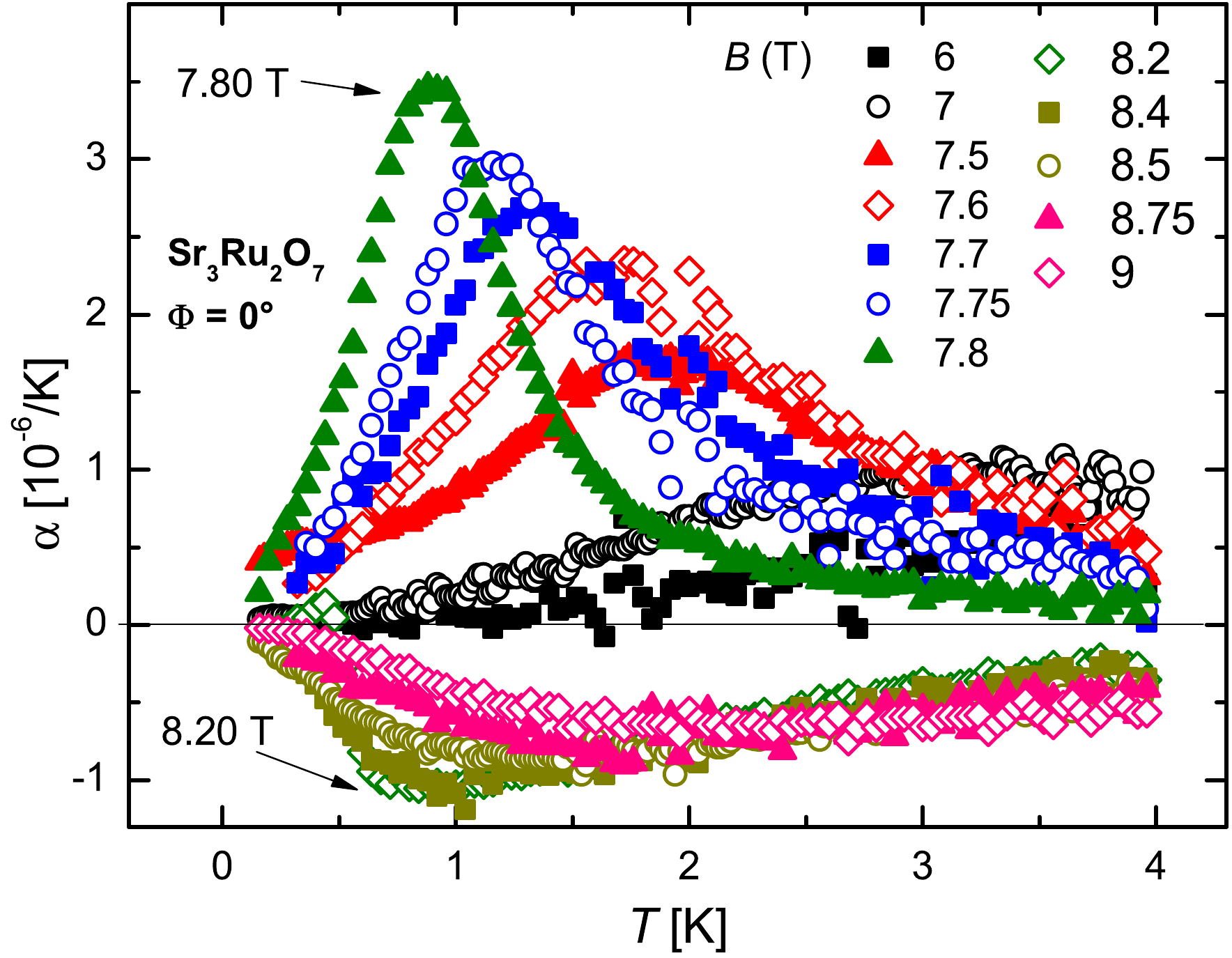}
	\caption{In-plane thermal expansion $\alpha$ for $B\parallel\text{c}$ after subtracting a Fermi liquid background of $0.14\cdot10^{-6}\,\text{K}^{-2}$. }
	\label{alpha_outside}
\end{figure}

\begin{figure}
\includegraphics[width=\columnwidth]{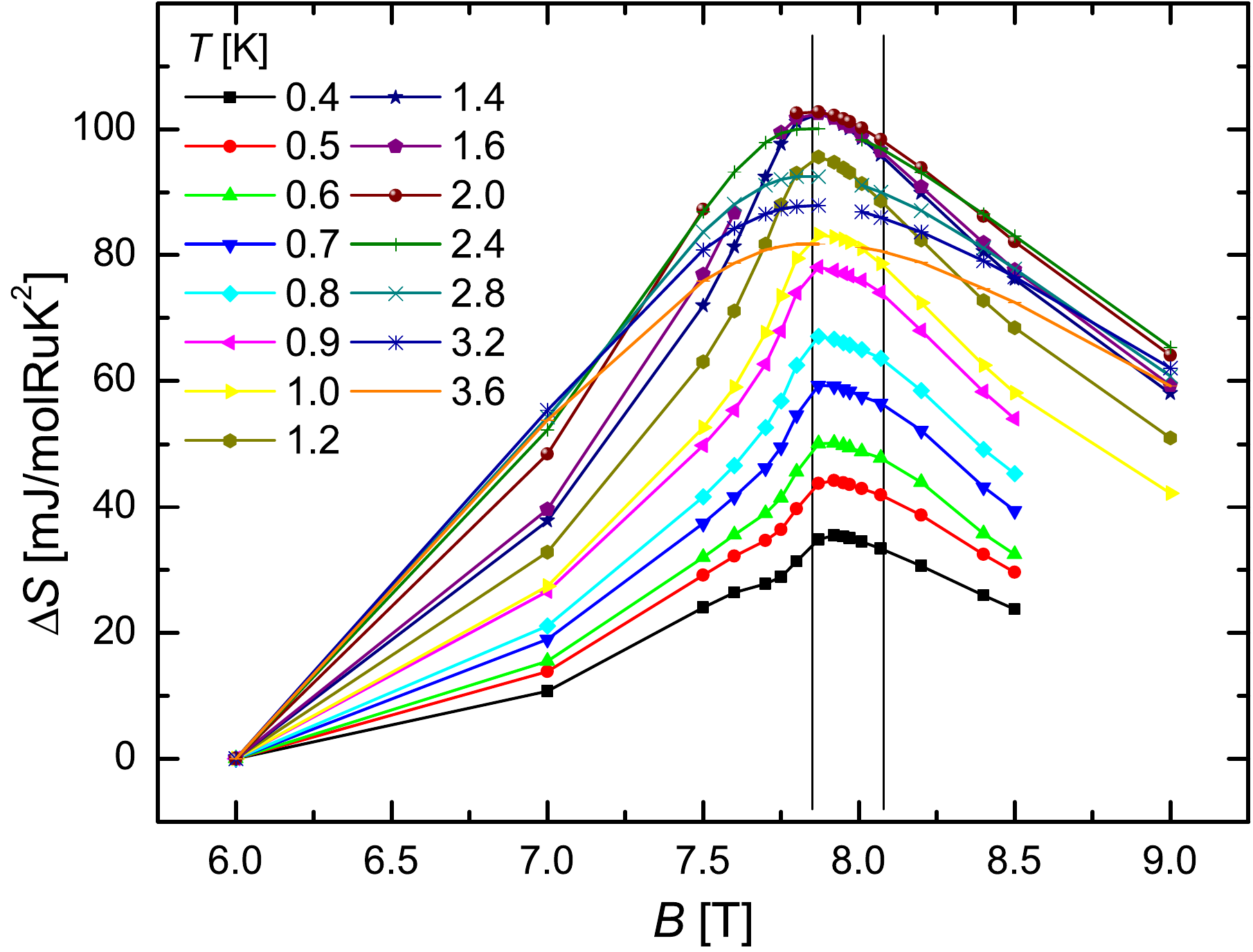}
\caption{Entropy of \system, calculated from the volume thermal expansion coefficient $\beta$ by using eq. (3). Vertical lines indicate $B_\text{c1}$ and $B_\text{c2}$.}
\label{S_vs_B}
\end{figure}

\begin{figure}
\includegraphics[width=\columnwidth]{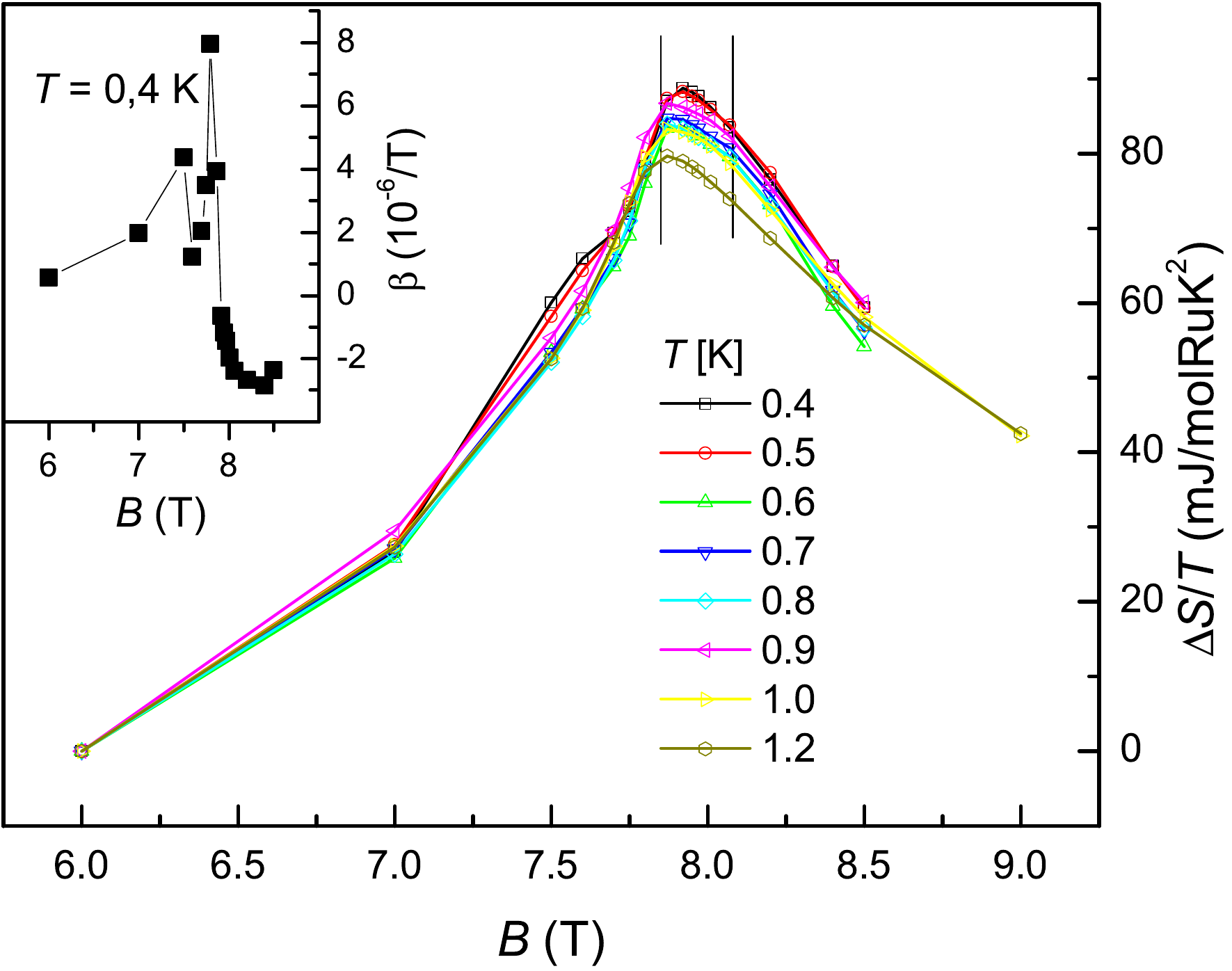}
\caption{Field dependence of the calculated entropy divided by temperature $S/T$, which is temperature-independent, as expected in the Fermi liquid regime. Vertical lines indicate $B_\text{c1}$ and $B_\text{c2}$. Inset: field-dependence of the volume thermal expansion coefficient $\beta$ at $T=0.4$~K.}
\label{SoverT_vs_B}
\end{figure}

In Fig.~\ref{alpha_outside}, we plot the in-plane thermal expansion coefficient for $B\parallel c$ at different magnetic fields. This can be combined with earlier data for the $c$-axis~\cite{gegenwart06-ms} in order to calculate the volume thermal expansion coefficient, $\beta=\alpha_c+2\cdot\alpha_{ab}$ for a tetragonal system. Since $|\alpha_c|>|\alpha_{ab}|$, $\beta$ (not shown) is dominated by the contribution of the $c$-axis thermal expansion, but shows in addition the nematic phase transition signatures which enter through the in-plane expansion.

Assuming that a unixial pressure enters the free energy only through a pressure-dependent critical field $H_\text{c}=H_\text{c}(p)$, the entropy change $\Delta S$ above $H_0=6$\,T can be calculated as
\begin{multline}
	\Delta S = S(H)-S(H_0) \\
	= V_\text{m}\left(\frac{\partial H_\text{c}}{\partial p}\right)^{-1}
		\int_{H_0}^H\beta(H')\text{d}H'
\end{multline}
with $\mu_0\partial H_\text{c}/\partial p = 5.6$\,T/GPa~\cite{gegenwart06,chiao02} and a molar volume $V_\text{m}=472$\,cm$^3$/mol\,Ru which can be calculated from the lattice parameters~\cite{shaked00}. Figure \ref{S_vs_B} shows that the entropy increases strongly on both the low- and high-field sides of the nematic phase as the critical field is approached. This is consistent with a divergence of the entropy in the approach of the quantum critical point which is cut off by the formation of a novel phase~\cite{rost09}.

In Fig.~\ref{SoverT_vs_B}, we plot the entropy divided by temperature for $T\leq 1.2$\,K. The data collapse well onto a single curve as is expected for a fermi liquid where $S$ is proportional to $T$. This also holds within the nematic phase which can therefore be characterized as a Fermi liquid, consistent with the observation of quantum oscillations~\cite{mercure2010}. At $B=7.5$\,T, we find $\Delta S/T\approx 50$\,mJ/molRuK$^2$, which is in very good agreement with the values determined by Rost et\,al.~\cite{rost09} from measurements of the magnetocaloric effect. This demonstrates the validity of the above scaling assumption between the field- and pressure derivatives of the entropy, which is characteristic for a pressure-dependent and field-tuned QCP.


\begin{figure}
\includegraphics[width=\columnwidth]{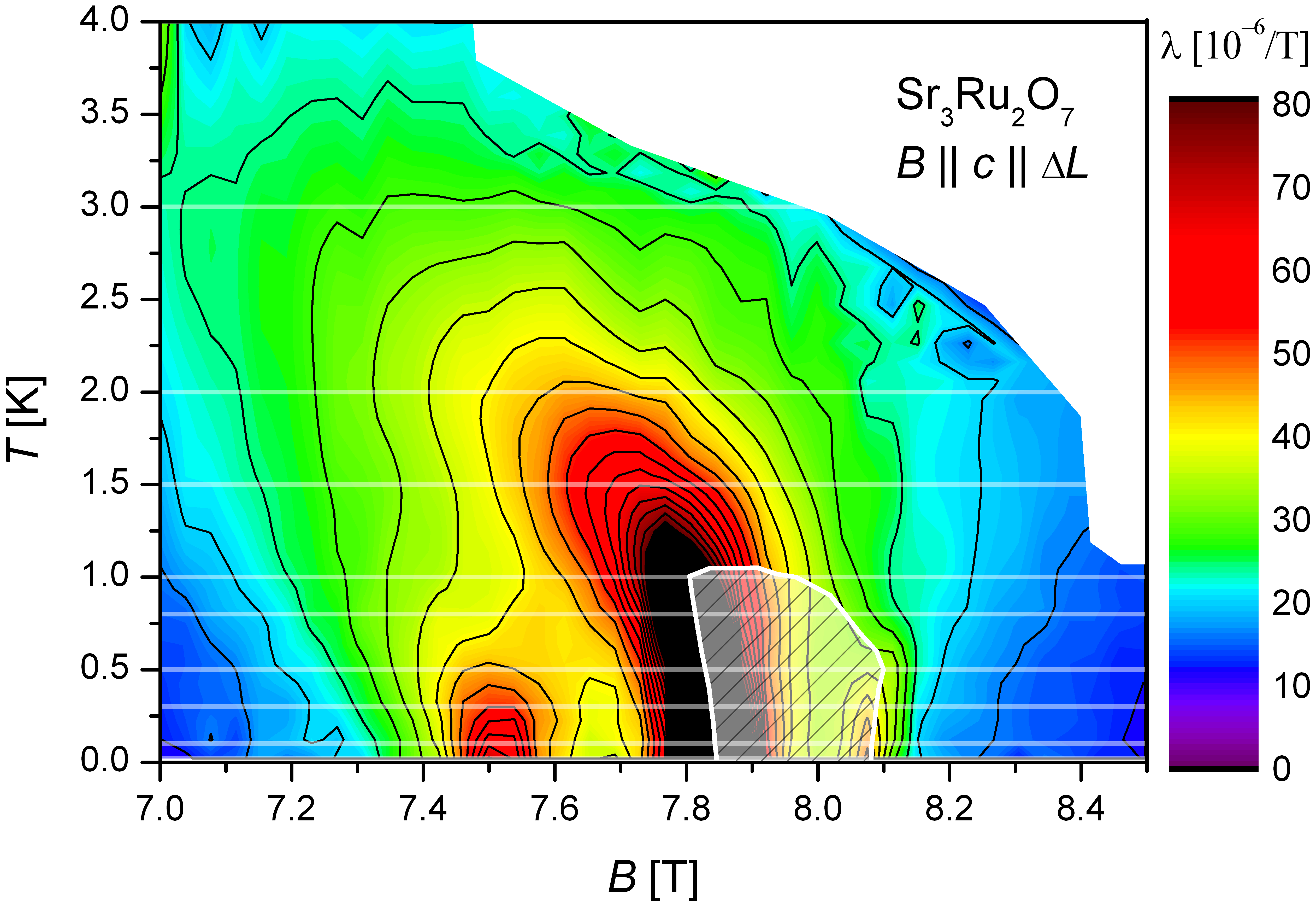}
\caption{Contours of the magnetostriction coefficient $\lambda$ for $B \parallel c \parallel \Delta L$. The hatched area indicates the nematic phase. The white horizontal lines indicate the temperatures at which magnetostriction data were taken~\cite{gegenwart06}.}
\label{lambda_c}
\end{figure}

The generic thermodynamic signatures of an itinerant metamagnetic quantum critical end point have recently been discussed by Weickert et\,al.~\cite{weickert10}. Within the scaling regime close to the QCP, it has been found that all second-order derivatives of the free energy display a similar divergence at the critical field. This results in a proportionality of the critical contributions to the differential suceptibility, magnetostriction and electronic compressibility, which all diverge in the approach of the QCP:

\begin{equation}
	\chi_\text{cr}\propto \lambda_\text{cr} \propto \kappa_\text{cr}
	\label{prop}
\end{equation}

The first proportionality in eq.~(\ref{prop}) is well fulfilled for \system~\cite{grigera04} and also for related systems like CeRu$_2$Si$_2$~\cite{flouquet02}. As $\chi\rightarrow\infty$ for $T\rightarrow 0$, the second proportionality predicts a divergence of the electronic compressibility. A 50\% reduction of the elastic modulus has indeed been found in CeRu$_2$Si$_2$~\cite{flouquet95}, although the quantum critical regime in this system is confined to temperatures above 0.5~K~\cite{weickert10}. For a generic system, one may expect that the metamagnetic QCP is preempted by a structural transition, due to the diverging behavior of the compressibility.

Previous thermal expansion measurements have revealed quantum critical scaling behavior in \system\ with a critical field of 7.845~T~\cite{gegenwart06}. Unfortunately, the elastic constants of \system\ have not been investigated in the approach of the critical field yet. However, we display in Fig.~\ref{lambda_c} a contour map of the isothermal magnetostriction, which under the assumption of quantum critical scaling is proportional to the electronic compressibility. Interestingly, at larger distances from the QCP, the shape of the contour lines resembles that of the nematic phase, particularly with the flanks tilted to the outside. This suggests a relation between quantum criticality and the nematic phase. However, the \enquote{roof} phase boundary with its transition temperature of approx. 1\,K is not parallel to a line of constant magnetostriction, but rather perpendicular. For the observed symmetry-breaking lattice distortion one would expect a softening of the in-plane shear modulus near this transition. A detailed investigation of the elastic properties along different directions will be important to study the relation between quantum critical fluctuations and the lattice distortion.

\section{Conclusion}

In this paper, we have studied the electronic nematic phase and its relation to quantum criticality in \system\ by high-resolution thermal expansion measurements. Upon entering the nematic phase at constant magnetic field, a second-order phase transition is detected in the in-plane thermal expansion. A fluctuation analysis would be compatible with a 2D Ising universality class, although the data are not conclusive, as the relative transition width is rather broad. As discussed previously~\cite{stingl11} the nematic phase is accompanied by domain formation. A preferential occupation of one of the two possible domain states can be achieved by a small in-plane field component resulting from a slight tilting angle $\Theta$ of the field with respect to the $c$-axis~\cite{borzi07}.

We have performed detailed measurements of the anisotropy of the in-plane thermal expansion at $\Theta\approx 5^\circ$ for various magnetic fields ranging from below $B_\text{c1}$ to above $B_\text{c2}$, where $B_\text{c1}$ and $B_\text{c2}$ denote two first-order metamagnetic transitions~\cite{grigera04}. For $B<B_\text{c1}$, we find that the system retains the four-fold rotational symmetry, while the latter is broken for fields in the nematic phase, $B_\text{c1}\leq B\leq B_\text{c2}$. Interestingly, we can detect a small thermal expansion anisotropy even at $B>B_\text{c2}$, i.e. outside the nematic phase. Measurements under larger tilted field angles can be found in~\cite{stingl11}, which reveal the in-plane distortion in the monodomain case.

We have also analyzed the volume thermal expansion and its quantum critical scaling. Assuming a proportionality between the pressure and field-derivatives, as expected in the close vicinity of a generic quantum critical end point~\cite{weickert10}, we have calculated the field-derivative of the entropy, which is found to nicely agree with direct measurements from~\cite{rost09}. We have also determined contours of the magnetostriction, which suggest a relation between quantum criticality and the nematic phase. Since metamagnetic quantum critical fluctuations generically soften the lattice, we speculate that the structural relaxation is driven by quantum criticality. However, a detailed study of the elastic constants is required to further characterize this interesting interplay.

We thank R. K\"{u}chler for his help in the construction of the
miniaturized dilatometer and M. Garst for helpful conversations. This work was supported by the Deutsche
Forschungsgemeinschaft within SFB~602.


\end{document}